\documentclass[aps,pra,twocolumn,showpacs,preprintnumbers,amsmath,amssymb]{revtex4}
\usepackage{graphicx}
\usepackage{bm}
\usepackage{dcolumn}

\def\prl#1#2#3{Phys. Rev. Lett. {\bf #1}, #2 (#3)}

\def\epjd#1#2#3{Eur. Phys. J. D {\bf #1}, #2 (#3)}
\def\pra#1#2#3{Phys. Rev. A {\bf #1}, #2 (#3)}

\def\epl#1#2#3{Europhys. Lett. {\bf #1}, #2 (#3)}
\def\njp#1#2#3{New J. Phys. {\bf #1}, #2 (#3)}
\def\jpamt#1#2#3{J. Phys. A: Math. Theor. {\bf #1}, #2 (#3)}
\def\jpamg#1#2#3{J. Phys. A: Math. Gen. {\bf #1}, #2 (#3)}
\def\pre#1#2#3{Phys. Rev. E {\bf #1}, #2 (#3)}
\def\pr#1#2#3{Phys. Rev. {\bf #1}, #2 (#3)}
\def\prb#1#2#3{Phys. Rev. B {\bf #1}, #2 (#3)}
\def\pknaw#1#2#3{Proc. K. Ned. Akad. Wet. {\bf #1}, #2 (#3)}

\def\noi{\noindent}
\def\bc{\begin{center}}
\def\ec{\end{center}}
 \newcommand{\bea}{\begin{equation}}
 \newcommand{\eea}{\end{equation}\noi}
 \newcommand{\ber}{\begin{eqnarray}}
 \newcommand{\eer}{\end{eqnarray}\noi}
\begin{document}
\title{Casimir force on interacting Bose-Einstein condensate}
\author{Shyamal Biswas$^1$}\email{tpsb2@iacs.res.in}
\author{J. K. Bhattacharjee$^{2}$}
\author{Dwipesh Majumder$^{1}$}
\author{Kush Saha$^{1}$}
\author{Nabajit Chakravarty$^{3}$}
%%%%%%%%%%%%%%%%%%
\vskip0.4cm \affiliation{$^1$Department of Theoretical Physics, Indian Association for the Cultivation of Science, Jadavpur, Kolkata 700032, India
 \vskip0.1cm
$^2$S.N. Bose National Centre for Basic Sciences, Sector 3, JD Block, Salt Lake, Kolkata 700098, India
\vskip0.1cm
$^3$Positional Astronomy Centre, Block AQ, Plot 8, Sector 5, Salt Lake, Kolkata 700091, India}

\date{\today}

\begin{abstract}
We have presented an analytic theory for the Casimir force on a Bose-Einstein condensate (BEC) which is confined between two parallel plates. We have considered Dirichlet boundary conditions for the condensate wave function as well as for the phonon field. We have shown that, the condensate wave function (which obeys the Gross-Pitaevskii equation) is responsible for the mean field part of Casimir force, which usually dominates over the quantum (fluctuations) part of the Casimir force.
\end{abstract}

\maketitle

\section{Introduction}
Confinement of the vacuum fluctuations of the electromagnetic field between two plates give rise to a long ranged Casimir force \cite{1}. The experimental verification \cite{2,3,4} of the Casimir effect is a confirmatory test of quantum field theory. Although the theory of Casimir force was first given \cite{1} for zero temperature ($T$), yet it can be generalized for any range of temperature \cite{5,6} and, for any dielectric substance \cite{7,8}  between two dielectric plates.  In general, Casimir forces are always present in nature when a medium with long ranged fluctuations is confined to restricted geometries \cite{9}. Consequently, it has been generalized for thermodynamical and critical systems \cite{9,10,11,12}.  Recent measurements \cite{13,14,15,16} of the Casimir forces for these kind of systems have drawn interests to the theoreticians \cite{9,12,17,18,19}. Although the Casimir force for a realistic BEC has not been measured, yet the Casimir-Polder force for the same was measured by the experimentalists of the Ref.\cite{20}. A successful theory for the Casimir-Polder force was also obtained by the authors of the Ref.\cite{21}. Since many new interesting phenomena are being discovered with the experimental studies \cite{22,23,24} of BEC, we are optimistic to have the measurement of the Casimir force on the BEC. On this issue, the Casimir effect on BEC has been the subject of a number of theoretical works within the last few years \cite{25,26,27,28,29,30,31,32}.

One of us recently studied \cite{28,29} the temperature dependence of the Casimir force for an ideal Bose gas below and above of its condensation temperature ($T_c$). Below the condensation point, it vanishes as a power law of the temperature and is long ranged. High above the condensation point, it vanishes exponentially and is short ranged. The role of geometry in the calculation of Casimir force for trapped ideal Bose gas was also explored in the Ref.\cite{29}. Different geometry causes different power law behavior \cite{28,29}. The Casimir force for the non-interacting BEC was also known to be vanished at the zero temperature \cite{28,29}. But, the Casimir force for interacting BEC at zero temperature may not vanish. The same force due to zero-temperature quantum fluctuations (phononic excitations) in an interacting one dimensional harmonically trapped (but homogeneous!) BEC was also obtained recently \cite{30}. It is to be mentioned, that, the theory of the Casimir force for a realistic (3D harmonically trapped and interacting) BEC has not been given so far.

We already have mentioned, that, the Casimir force depends on the geometry of the system. From this dependence, we can expect, that, the different geometry of the same system may not change the basic nature of the Casimir force. Hence, instead of the studying the Casimir force for a realistic BEC, it is relevant to study the Casimir force for a 3D interacting BEC in the plate geometry. On this issue, the Casimir effect due to the zero-temperature quantum fluctuations (phononic excitations) of a homogeneous weakly-interacting dilute BEC confined to a parallel plate geometry, has recently been studied with periodic boundary conditions \cite{27,32}.

Within the plate geometry, the BEC is considered to be homogeneous when the plate separation is infinitely large. Otherwise, the condensate is inhomogeneous, and it obeys Dirichlet boundary conditions. Since a macroscopic number of particles below $T_c$ occupy the single particle ground state, the theoretical predictions of the BEC for $T \rightarrow 0$, match well with the experimental results which are actually obtained for $T\lnsim T_c$. Hence, we are interested in calculating the Casimir force for $T\rightarrow 0$ on a 3D interacting inhomogeneous BEC confined between two parallel plates.

Our calculation starts from the standard grand canonical Hamiltonian ($\hat{\mathcal H}$) for the interacting Bose particles. Then we obtain the grand potential for the condensate, and quadratic (quantum) fluctuations for the phonon field. The grand potential is the mean field part of the $\hat{\mathcal H}$. The condensate wave function obeys the Gross-Pitaevskii (G-P) equation \cite{33}. From the consideration of the Dirichlet boundary conditions, the condensate becomes inhomogeneous. Then we obtain the mean field force acting on the plates in terms of the maximum of the inhomogeneous condensate wave function.  The maximum of the condensate wave function is asymptotically obtained for very small and very large limits of the plate separation. For the whole range of the plate separation, we obtain an interpolation formula, which matches very well with the asymptotic solutions. We also justify this interpolation from the exact graphical solutions. We obtain the mean field Casimir force from the interpolation formula. Then we obtain the Casimir force for the vacuum fluctuations of the phonon field with the Dirichlet boundary conditions. Finally, we compare the mean field and the quantum fluctuations contributions to the Casimir force.

\section{Elementary excitations over a BEC in the plate geometry}
Let us consider a Bose gas of $N$ identical particles to be confined between two infinitely large parallel plates of area $A$ along the $x-y$ plane and their separation along the $z$ direction be $L$. For $T\rightarrow 0$, almost all the particles ($N_0\approx N$) form the condensate, and the elementary excitations over the condensate be perturbatively treated. The elementary excitations over an unbounded condensate are obtained by the standard textbook approach \cite{annett,33}. We follow the same approach for obtaining the elementary excitations over a bounded condensate in the plate geometry.

Let the position vector and mass of a single Bose particle be denoted as ${\bf r}={\bf r}_\perp+z\hat{k}$ and $m$ respectively. The grand canonical Hamiltonian operator for such a system of interacting Bose gas is given by \cite{33}
\begin{eqnarray}
\hat{\mathcal H}&=&\int\hat{\Psi}^{\dagger}({\bf r})\bigg(-\frac{\hbar^2}{2m}\nabla^2-\mu\bigg)\hat{\Psi}({\bf r})d^3{\bf r}\nonumber \\ &+&\frac{1}{2}\int\int\bigg(\hat{\Psi}^{\dagger}({\bf r})\hat{\Psi}^{\dagger}({\bf r'})V({\bf r-r'})\hat{\Psi}({\bf r})\hat{\Psi}({\bf r'})\bigg)d^3{\bf r}d^3{\bf r'},\nonumber\\
\end{eqnarray}
where $\mu$ is the chemical potential, $V({\bf r-r'})$ is the inter-particle interaction potential, and $\hat{\Psi}({\bf r})$ is the field operator for the Bose particles. For the simplest case, the interaction potential can be considered as $V({\bf r-r'})=g\delta^3({\bf r-r'})$, where $g$ is the coupling constant. The connection of this coupling constant with the s-wave scattering length ($a_s$) is \cite{33} $g=\frac{4\pi\hbar^2a_s}{m}$. The field operator in terms of the single particle orthonormal wave functions $\{\phi_i({\bf r})\}$ is expressed as $\hat{\Psi}({\bf r})=\sum_{i=0}^{\infty}\phi_i({\bf r})\hat{a}_i$, where $\hat{a}_i$ and $\hat{a}_i^{\dagger}$ annihilates and creates respectively a Bose particle at the state $\phi_i({\bf r})$. Within the perturbative approach, the grand canonical Hamiltonian in terms of the excitations $\delta\hat{\Psi}({\bf r})$ ($=\hat{\Psi}({\bf r})-\sqrt{N_0}\phi_0({\bf r})$) be effectively written up to the quadratic order as \cite{annett}
\begin{eqnarray}
\hat{\mathcal H}=\Omega_0&+&\int\delta\hat{\Psi}^{\dagger}({\bf r})\bigg(-\frac{\hbar^2}{2m}\nabla^2\bigg)\delta\hat{\Psi}({\bf r})d^3{\bf r}\nonumber \\ &+&\frac{gn}{2}\int\bigg(2\delta\hat{\Psi}^{\dagger}({\bf r})\delta\hat{\Psi}({\bf r})+\delta\hat{\Psi}^{\dagger}({\bf r})\delta\hat{\Psi}^{\dagger}({\bf r})\nonumber \\ &+& \delta\hat{\Psi}({\bf r})\delta\hat{\Psi}({\bf r})\bigg) d^3{\bf r},
\end{eqnarray}
where 
\begin{eqnarray}
\Omega_0=N_0\int\bigg(\frac{\hbar^2}{2m}|\nabla\phi_0({\bf r})|^2-\mu|\phi_0({\bf r})|^2+\frac{gN_0}{2}|\phi_0({\bf r})|^4\bigg)d^3{\bf r}\nonumber\\
\end{eqnarray}
is the grand potential for the condensate, and where $N_0\phi_0^2({\bf r})$ terms and $gn$ in the quadratic fluctuations are approximated by the bulk density ($n$) and the chemical potential ($\mu$) respectively \cite{annett}. Within the perturbative approach we can write the quantum fluctuations for the Dirichlet boundary conditions as $\delta\hat{\Psi}({\bf r})=\sqrt{\frac{2}{LA}}\sum_{j=1}^{\infty}\int\text{sin}\big(\frac{j\pi z}{L}\big)e^{i\frac{{\bf p}_{\perp}.{\bf r}_{\perp}}{\hbar}}\hat{a}_{{\bf p}_{\perp},j}\frac{Ad^2{\bf p}_{\perp}}{(2\pi\hbar)^2}$, where $\hat{a}_{{\bf p}_{\perp},j}$ annihilates a Bose particle of $x-y$ momentum ${\bf p}_{\perp}=p_x\hat{i}+p_y\hat{j}$ and energy $\frac{p_\perp^2}{2m}+\frac{\pi^2\hbar^2j^2}{2mL^2}=\frac{p^2}{2m}$. The $\hat{\mathcal H}$ in the Eqn.(2) can now be diagonalized in terms of the phononic operators through the Bogoliubov transformations \cite{33} $\hat{a}_{{\bf p}_{\perp},j}=u_{p_\perp,j}\hat{b}_{{\bf p}_{\perp},j}+v_{p_\perp,j}\hat{b}_{{\bf -p}_{\perp},j}^{\dagger}$ and $\hat{a}_{{\bf p}_{\perp},j}^{\dagger}=u_{p_\perp,j}\hat{b}_{{\bf p}_{\perp},j}^{\dagger}+v_{p_\perp,j}\hat{b}_{{\bf -p}_{\perp},j}$, where \cite{33} $u_{p_\perp,j}=\big(\frac{p^2/2m+gn}{2\epsilon(p_\perp,j)}+\frac{1}{2}\big)^{1/2}$, $v_{p_\perp,j}=-\big(\frac{p^2/2m+gn}{2\epsilon(p_\perp,j)}-\frac{1}{2}\big)^{1/2}$, and \cite{33,roberts}
\begin{eqnarray}
\epsilon(p_\perp,j)=\bigg(\frac{gn}{m}\bigg(p_\perp^2+\frac{\pi^2\hbar^2 j^2}{L^2}\bigg)\bigg(1+\frac{p_\perp^2+\frac{\pi^2\hbar^2 j^2}{L^2}}{4mgn}\bigg)\bigg)^{1/2}.\nonumber\\
\end{eqnarray}
With the above transformations, we recast the Eqn.(2) in terms of the phononic excitations as
\begin{eqnarray}
\hat{\mathcal H}=\Omega_0&+&\frac{1}{2}\sum_{{\bf p}_{\perp},j}\bigg(\epsilon(p_\perp,j)-\big(\frac{p_\perp^2}{2m}+\frac{\pi^2\hbar^2j^2}{2mL^2}\big)-gn\bigg)\nonumber\\&+&\sum_{{\bf p}_{\perp},j}\epsilon(p_\perp,j)\hat{b}_{{\bf p}_{\perp},j}^{\dagger}\hat{b}_{{\bf p}_{\perp},j}.
\end{eqnarray}
For $T\rightarrow 0$, there would be no phonon, and consequently, the grand canonical energy ($<\hat{\mathcal H}>$) of the system for the vacuum of the phonon can be obtained from the Eqn.(5) as
\begin{eqnarray}
{\mathcal E}_0=\Omega_0+\varepsilon_0(L,n),
\end{eqnarray}
where $\varepsilon_0(L,n)=\frac{1}{2}\sum_{{\bf p}_{\perp},j}\big(\epsilon(p_\perp,j)-\big(\frac{p_\perp^2}{2m}+\frac{\pi^2\hbar^2j^2}{2mL^2}\big)-gn\big)$ is the contribution to the grand potential (${\mathcal E}_0$) due to the quantum (vacuum) fluctuations of the phonon field. Roberts and Pomeau \cite{roberts} also correctly predicted this term from the original work of Lee, Huang and Yang \cite{37} on a homogeneous condensate.

From the second term of the Eqn.(6), we will obtain an expression for the Casimir force due to the vacuum fluctuations. Irrespective of the form of the second term, our expectation for the Casimir force due to the vacuum fluctuations of the phonon field, would be similar to that due to the vacuum fluctuations of photon (electromagnetic) field. Since, the Casimir force for the vacuum fluctuations of photon field is $C_{ptn}=-\frac{A\pi^2\hbar c}{240L^4}$ \cite{1}, our expectation for that due to vacuum fluctuations of the phonon field ($\delta\hat{\Psi}$) would primarily be \cite{25,roberts}
\begin{eqnarray}
C_{phn}=-\frac{A\pi^2\hbar v(n)}{480L^4},
\end{eqnarray}
where the speed ($c$) of light is replaced by the speed ($v(n)=\sqrt{gn/m}$) of phonon, and a factor 1/2 appears for a single polarization of a phonon in a BEC.

Before going into the details of the quantum fluctuations part, let us calculate the mean field part of the Casimir force from the first term of the Eqn.(6).
\section{Mean field force acting on the plates}
In the first term ($\Omega_0$) of the Eqn.(6), we can replace $\phi_0({\bf r})$ by $\phi_0(z)$ due to the fact, that, the condensate is placed between two infinitely large parallel plates, and the motion of the condensate is relevant only in the $z$ direction. This term governs the equation of motion ($\frac{\delta\Omega_0}{\delta\phi_0^*(z)}=0$) for the BEC as \cite{33}
\begin{eqnarray}
\bigg(-\frac{\hbar^2}{2m}\frac{d^2}{dz^2}-\mu+gN_0|\phi_0(z)|^2\bigg)\phi_0(z)=0,
\end{eqnarray}
which is known as time independent Gross-Pitaevskii (G-P) equation. In the above analysis, $\phi_0(z)$ has been considered to be a complex function which can obviously be written in terms of the condensate density ($n_0(z)$) and a real phase ($\theta(z)$) as $\phi_0(z)=\sqrt{\frac{n_0(z)}{N_0}}e^{i\theta(z)}$.

We are considering the fact that no macroscopic part of the condensate is moving as a whole within the plate geometry. This consideration allows us to write ${\bf J}\rightarrow 0$, where ${\bf J}=n(z)\frac{\hbar}{m}\nabla\theta(z)$ is the particle current density. If ${\bf J}\rightarrow 0$, we can ignore the phase part of the kinetic energy density ($\frac{\hbar^2}{2m}[(\nabla\sqrt{n_0(z)})^2+n_0(z)(\nabla\theta(z))^2]$), and can write $\frac{\nabla\theta(z)}{\big(\frac{1}{\sqrt{n_0(z)}}\big)\big(\nabla\sqrt{n_0(z)}\big)}\rightarrow 0$ which of course allows the no current case to be applied. Under the same condition, we consider $\phi_0(z)$ to be a real function for the rest of this paper.

Now, with the substitutions $a=2\mu m/\hbar^2$ and $b=2mgN_0/\hbar^2$, the Eqns.(3) and (8) can be recast as
\begin{eqnarray}
\Omega_0=&\frac{N_0A\hbar^2}{m}\int_0^L(\frac{1}{2}(\frac{d\phi_0(z)}{dz})^2-\frac{a}{2}\phi^2_0(z)+\frac{b}{4}\phi^4_0(z))dz\nonumber\\
\end{eqnarray}
and
\begin{eqnarray}
-\frac{d^2\phi_0(z)}{dz^2}-a\phi_0(z)+b\phi^3_0(z)=0
\end{eqnarray}
respectively.

From the Eqn.(9), we get the mean field force acting on the plates as $F_{mf}(L)=-\frac{\delta\Omega_0}{\delta L}\big|_{a}=-\frac{\partial\Omega_0}{\partial L}\big|_{a}-\frac{NA\hbar^2}{m}(\frac{1}{2}(\frac{d\phi_0}{dz})^2\big|_{z=L}-\frac{a}{2}\phi_0^2(L)+\frac{b}{4}\phi_0^4(L))$, where $\frac{\partial\Omega_0}{\partial L}\big|_a=\frac{NA\hbar^2}{m}\int_0^L(\frac{d\phi_0}{dz}\frac{d}{dL}(\frac{d\phi_0}{dz})-a\phi_0\frac{d\phi_0}{dL}+b\phi_0^3\frac{d\phi_0}{dL})dz$ is associated with the change in the wave function with respect to the change in the plate separation, and the other part of $\frac{\delta\Omega_0}{\delta L}\big|_{a}$ is associated with the change in the integration limit. Now, from the Dirichlet boundary conditions ($\phi_0(0)=\phi_0(L)=0$) and from the Eqn.(10), we recast the above expression of the mean field force as
\begin{eqnarray}
F_{mf}(L)=-\frac{NA\hbar^2}{m}\bigg(\frac{d\phi_0}{dz}\frac{d\phi_0}{dL}\bigg|_0^L+\frac{1}{2}\bigg(\frac{d\phi_0}{dz}\bigg)^2\bigg|_{z=L}\bigg).
\end{eqnarray}
Since $\phi_0(z)$ obeys Dirichlet boundary conditions, we can write it in terms of the Fourier modes as $\phi_0(z)=\sum_{j=1}^{\infty}f_1(j)\times\text{sin}(\frac{j\pi z}{L})$, where $f_1(j)$ is to be determined from the Eqn.(10). However, we will calculate the mean field force without determining $f_1(j)$. From the Fourier expansion, it is easy to check that $\frac{d\phi_0}{dL}\big|_{z=0}=0$ and $\frac{d\phi_0}{dL}\big|_{z=L}=-\frac{d\phi_0}{dz}\big|_{z=L}$. With these properties, the Eqn.(11) becomes \cite{19}
\begin{eqnarray}
F_{mf}(L)=\frac{NA\hbar^2}{m}\frac{1}{2}\bigg(\frac{d\phi_0}{dz}\bigg)^2\bigg|_{z=L}.
\end{eqnarray}
This is to be mentioned that, the Eqn.(12) can also be obtained from the stress tensor \cite{durand} of the G-P equation. In the Ref.\cite{durand}, the stress tensor was calculated for non-interacting bosons and fermions. The stress tensor (whose average over the statistical distribution is actually the pressure at any point of the system) involves all possible orthonormal single particle states and their occupations. If the inter-particle interaction be introduced in the formalism of the Ref.\cite{durand}, the Schrodinger equation for the ground state will become the G-P equation. With these considerations, the stress tensor of the Ref.\cite{durand} will give exactly the same force at the boundary ($z=L$) of the system as we have obtained in the Eqn.(12).

Besides the Dirichlet boundary conditions $\phi_0(0)=\phi_0(L)=0$, $\phi_0(z)$ is obviously symmetric about $z=\frac{L}{2}$, and, since $\phi_0(z)$ is the ground state of the system, it must have a single maximum at $z=\frac{L}{2}$. Let us now multiply the Eqn.(10) by $2\frac{d\phi_0}{dz}$ and integrate with respect to $z$ from $z=\frac{L}{2}$ to $z=z$. From this integration and with the fact that $\frac{d\phi_0}{dz}|_{z=\frac{L}{2}}=0$, we have
 \bea
\bigg(\frac{d\phi_0(z)}{dz}\bigg)^2=a\bigg(\phi_0^2\bigg(\frac{L}{2}\bigg)-\phi_0^2(z)\bigg)+\frac{b}{2}\bigg(\phi_0^4(z)-\phi_0^4\bigg(\frac{L}{2}\bigg)\bigg),
 \eea
where $\phi_0^2(\frac{L}{2})$ is an integrating constant. From the Eqn.(13) and from the Dirichlet boundary conditions, we recast the Eqn.(12) as \cite{19}
\begin{eqnarray}
F_{mf}(L)&=&\frac{NA\hbar^2}{m}\frac{1}{2}\bigg(a\phi_0^2\bigg(\frac{L}{2}\bigg)-\frac{b}{2}\phi_0^4\bigg(\frac{L}{2}\bigg)\bigg)\nonumber\\&=&\frac{NA\hbar^2a^2}{bm}\eta(1-\eta),
\end{eqnarray} 
where $\eta=\frac{b}{2a}\phi_0^2(\frac{L}{2})$. From the Eqn.(14) we see that, we need only to know $\eta$ as well as the maximum of $\phi_0(z)$ for obtaining the mean field force acting on the plates. In the next section we will write a transcendental equation for the maximum of $\phi_0(z)$.

\section{Analysis of the Gross-Pitaevskii equation}

Eqn.(13) can once again be written in the differential form
\begin{eqnarray}
dz=\frac{d\phi_0(z)}{\sqrt{a(\phi_0^2(\frac{L}{2})-\phi_0^2(z))+\frac{b}{2}(\phi_0^4(z)-\phi_0^4(\frac{L}{2}))}}.
\end{eqnarray}
Now, integrating the Eqn.(15) from $z=0$ to $z=z$, we get
\begin{eqnarray}
z=\frac{1}{\sqrt{a(1-\eta)}}\int_0^{\frac{\phi_0(z)}{\phi_0(\frac{L}{2})}}\frac{dt}{\sqrt{(1-t^2)(1-Mt^2)}}
\end{eqnarray}
where $t=\frac{\phi_0(z)}{\phi_0(\frac{L}{2})}$ and $M=\frac{\eta}{1-\eta}$ \cite{carr}. From the Eqn.(16), we get the ground state \cite{carr}
\begin{eqnarray}
\phi_0(z)&=&\frac{\sqrt{2M}}{L\sqrt{b}}2 \text{EllipticK}[M]\nonumber\\&\times&\text{JacobiSN}[2 \text{EllipticK}[M]\frac{z}{L},M],
\end{eqnarray}
where $\text{EllipticK}[M]=\frac{\pi}{2}[1+\frac{M}{4}+\frac{9}{64}M^2+\frac{25}{256}M^3+...]$ is the complete elliptic integral of first kind, and $\text{JacobiSN}[x,y]$ is a Jacobian elliptic function. Eqn.(17) represents an exact solution of the Eqn.(15). The normalization condition $A\int_0^L\phi^2_0(z)dz=1$ can be recast from the Eqn.(17) as \cite{carr}
 \bea
\frac{8 \text{EllipticK}[M]}{b}\bigg(\text{EllipticK}[M]-\text{EllipticE}[M]\bigg)=\frac{L}{A},
 \eea
where $\text{EllipticE}[M]=\frac{\pi}{2}[1-\frac{M}{4}-\frac{3}{64}M^2-\frac{5}{256}M^3-...]$ \cite{34} is also another elliptic integral. Now, putting $z=L/2$ and $M=\frac{\eta}{1-\eta}$ in the Eqn.(17), we get \cite{19}
\begin{eqnarray}
\frac{L}{2}=\frac{1}{\sqrt{a}}\frac{\text{EllipticK}[\frac{\eta}{1-\eta}]}{\sqrt{1-\eta}},
\end{eqnarray}
which can once again be recast in the form of the ground state energy \cite{carr}
 \bea
\mu=\frac{2\hbar^2\text{EllipticK}^2[M](1+M)}{mL^2}.
 \eea 
Eqn.(19) is a transcendental equation for the maximum of $\phi_0(z)$. $M$ can be determined form the transcendental Eqn.(18) by a graphical method \cite{carr}. Once we put the determined $M$ in to the Eqn.(20), we will get the ground state energy.

It is easy to check from the Eqn.(19) and (20) that, the allowed ranges of $\eta$ and $M$ are $0\le\eta<\frac{1}{2}$ and $0\le M<1$ respectively. In the non-interacting limit ($M\rightarrow 0$), Eqn.(20) gives the ground state energy of a particle in a box as $\mu=\frac{\pi^2\hbar^2}{2mL^2}$. On the other hand, in the bulk limit ($M\rightarrow 1$), $\frac{\text{EllipticE}[M]}{\text{EllipticK}[M]}\rightarrow 0$ can be put in the Eqn.(18), which together with the Eqn.(20) gives \cite{33} $\mu=gn$, which we already have used in the Eqn.(2).

Since the equilibrium of our system is obtained by minimizing the grand potential, it is relevant to get the force (acting on the plates), from the derivative of the grand potential with respect to the plate separation. For this process, it is necessary to keep the chemical potential unchanged. To get the Casimir force, one needs to subtract the bulk force ($F_{mf}(\infty)$) from the actual force ($F_{mf}(L)$) in the Eqn.(14) with constant chemical potential, which can conveniently be taken as the bulk chemical potential $\mu=gn$ even for this finite system \cite{pitaevskii}.

Let us now look at the Eqn.(14), where $\frac{N}{b}$ ($=\frac{\hbar^2}{2mg}$ by the definition is a constant, and $a$ ($=2\mu m/\hbar^2$) is to be kept constant ($2gnm/\hbar^2$) for the measurement of the $F_{mf}(L)$, and that of the Casimir force. Therefore, for obtaining the Casimir force, we need only to know the $\eta$ (in the Eqn.(14)) in terms of $a$ and $L$. From the Eqn.(19), we can determine $\eta$ in terms of $a$ and $L$ by asymptotic and graphical analyses.

\section{Asymptotic and graphical analyses of $\eta$}
\subsection{Asymptotic analysis}
\begin{figure}
\includegraphics{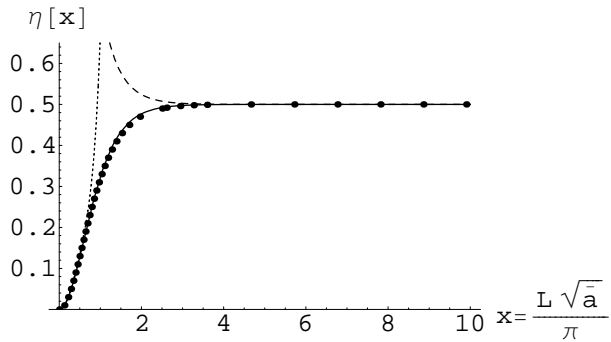}
\caption {Comparison between the interpolated $\eta$ (continuous line), the asymptotic $\eta$'s of the extreme limits, and the graphical solutions (dots) of $\eta$ . The dots are obtained from the graphical solutions of the Eqn.(24) with the help of Wolfram Mathematica 5.2. The dotted, dashed and the continuous lines follow the Eqns.(21), (22) and (23) respectively.}
\label{fig:}
\end{figure}

It is easy to check from the Eqn.(19) that, for $\eta\rightarrow 0$, $a\rightarrow \frac{\pi^2}{L^2}$ and for $\eta\rightarrow \frac{1}{2}$, $aL^2\rightarrow\infty$. Let us define $\bar{a}=a-\frac{\pi^2}{L^2}$. Now we can check from the Eqn.(19) that, $\eta$ as well as $\phi_0(\frac{L}{2})$ goes to zero as $L\sqrt{\bar{a}}\rightarrow 0$. For $\eta\rightarrow 0$, we can expand the right hand side of Eqn.(19) up to the third order in $\eta$ and get
\begin{eqnarray}
\eta&=&\frac{2}{3}\bigg(\frac{L\sqrt{\bar{a}}}{\pi}\bigg)^2\bigg(1-\frac{25}{24}\bigg(\frac{L\sqrt{\bar{a}}}{\pi}\bigg)^2+1.04514\bigg(\frac{L\sqrt{\bar{a}}}{\pi}\bigg)^4\nonumber\\ &&+...\bigg)  \ \ \text{for}  \ \  \frac{L\sqrt{\bar{a}}}{\pi}\ll 1.
\end{eqnarray}

In the other limit, i.e. for $\eta\rightarrow\frac{1}{2}$, $L\sqrt{\bar{a}}$ goes to infinity. In this limit $\eta$ can be expressed as $\eta=\frac{1}{2}(1-\delta)$, where $\delta\rightarrow 0$. Let us now evaluate the $\delta$ from the Eqn.(19), which can be expressed in the integral form with the substitution $p=1-t$ as $\frac{L\sqrt{\bar{a}+\frac{\pi^2}{L^2}}}{2}=\frac{1}{\sqrt{\frac{1+\delta}{2}}}\int_0^1\frac{dp}{\sqrt{p\big(2-p\big)\big(1-\frac{1-\delta}{1+\delta}(1-2p+p^2)\big)}}$, which says that $p=0$ has a logarithmic divergence, so that most of the integral would come from $p\rightarrow 0$. In this limit the above integral can be approximated with the first order in $\delta$ and $p$, as $\frac{L\sqrt{\bar{a}}}{2}=\int_0^1\frac{dp}{\sqrt{2p(\delta+p)}}=\sqrt{2}\text{sinh}^{-1}\big(\frac{1}{\sqrt{\delta}}\big)$, which asymptotically gives
\begin{eqnarray}
\eta=\frac{1}{2}\text{coth}^2\bigg(\frac{L\sqrt{\bar{a}/2}}{2}\bigg) \ \ \text{for}  \ \  \frac{L\sqrt{\bar{a}}}{\pi}\gg 1.
\end{eqnarray}
Let us now find a smooth function for $\eta$ as an interpolation for the whole range $0\le\eta<\frac{1}{2}$, in such a way, that, it fits to the extreme ends of $\eta$. Since $\phi_0(z)=\sqrt{\frac{a}{b}}\text{tanh}(z\sqrt{a/2})$ is a solution of the Eqn.(10) for an unbounded situation, we can take a trial for the bounded system that, $\phi_0(\frac{L}{2})$ would be close to $\sqrt{\frac{a}{b}}\text{tanh}\big(\frac{L\sqrt{\bar{a}/2}}{2}\big)$. With this consideration, we can take a trial interpolation as \cite{bhattacharyya}
\begin{eqnarray}
\eta=\frac{1}{2}\text{tanh}^2\bigg(\frac{L\sqrt{\bar{a}/2}}{2}\bigg) \ \ \text{for}  \ \   0\le\frac{L\sqrt{\bar{a}}}{\pi}<\infty.
\end{eqnarray} 
For $\frac{L\sqrt{\bar{a}}}{\pi}\gg 1$, the trial interpolation in the Eqn.(23), fits obviously well with the $\eta$ of the Eqn.(22). For $\frac{L\sqrt{\bar{a}}}{\pi}\ll 1$, the trial interpolation fits also well with the $\eta$ of the Eqn.(21). We can see how well the interpolation fits with the two asymptotic $\eta$s in the FIG. 1. Although the asymptotic analysis is easier to express mathematically, yet the interpolation is not beyond the doubt for the region $0.5 \lesssim x\lesssim 3$ in the FIG 1. For this reason, we need the graphical analysis to support the interpolation formula.
\subsection{Graphical analysis}

\begin{figure}
\includegraphics{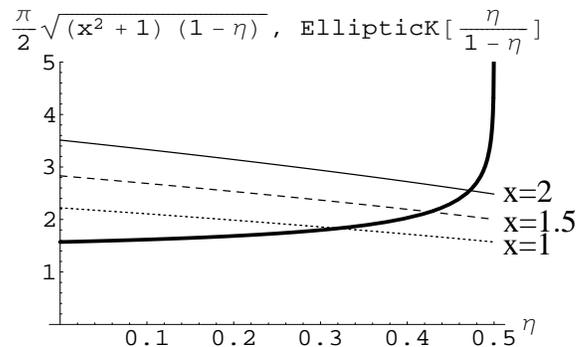}
\caption {Thick continuous line represent the right hand side of the Eqn.(24). The other lines represent the left hand side of the Eqn.(24) for $x=\frac{L\sqrt{\bar{a}}}{\pi}=1.0, 1.5, \text{and}\ 2$. Intersecting points give the graphical solutions $\eta=0.3206, 0.4264, \text{and}\ 0.4721$ respectively. Along with other such points, we plot the dots of the FIG. 1.}
\label{fig:}
\end{figure}

Eqn.(19) can once again be recast in terms of $x=\frac{L\sqrt{\bar{a}}}{\pi}$ as
 \bea
\frac{\pi}{2}\sqrt{x^2+1}\sqrt{1-\eta}=\text{EllipticK}\bigg[\frac{\eta}{1-\eta}\bigg].
 \eea
For a given $x$, we plot the left hand side of the Eqn.(24) with respect to $\eta$ in the FIG. 2. In the same figure, we also plot the right hand side of the Eqn.(24) with respect to $\eta$. The intersection point is the graphical solution of $\eta$ for the given $x$. For a given set $\{x\}$, we can get the corresponding graphical solution $\{\eta\}$. We plot this set of points in the FIG. 1, and can compare the interpolation formula (Eqn.(23)) with the graphical solutions of $\eta$.

Now, we see in the FIG. 1 that, for the entire region $0\le x\lesssim 10$, the interpolation fits very well with the graphical solution of $\eta$. So, for the rests of this paper, we will calculate the Casimir force from our interpolation formula (Eqn.(23)).

\section{Casimir force}
\subsection{Mean field part of the Casimir force on a BEC}

Since in the bulk limit $\eta\rightarrow\frac{1}{2}$, we get the bulk force from the Eqn.(14) as $F_{mf}(\infty)=\frac{NA\hbar^2a^2}{4mb}$. Subtracting this bulk force from the mean field force in the Eqn.(14), we get the mean field Casimir force
\begin{eqnarray}
C_{mf}(L,n)=-\frac{NA\hbar^2a^2}{mb}\bigg(\frac{1}{4}-\eta(1-\eta)\bigg),
\end{eqnarray}
which is obviously attractive, and can be simplified with the Eqn.(23) in the following form
\begin{eqnarray}
C_{mf}(L,n)=-\frac{NA\hbar^2a^2}{4mb}\text{sech}^4\bigg(\sqrt{\frac{L^2a-\pi^2}{8}}\bigg).
\end{eqnarray}
Now, putting the values of the constants $a$, $b$ and $g$, we recast the Eqn.(26) as
\begin{eqnarray}
C_{mf}(L,n)=-\frac{A2\pi\hbar^2 a_s n^2}{m}\text{sech}^4\bigg(\sqrt{\frac{8\pi L^2 a_s n-\pi^2}{8}}\bigg).
\end{eqnarray} 
It is clear from the Eqn.(27) that, for $a_s>0$, there exists a lower critical density $n_c=\frac{\pi}{8L^2 a_s}$ of atoms, at and below which we can not write $\mu=gn$. The critical density also defines a critical phonon speed $v_c=\sqrt{gn_c/m}$. Now, the Eqn.(27) can be expressed in terms of the unit less `density' $\rho=\frac{n}{n_c}$ as
\begin{eqnarray}
C_{mf}(L,\rho)&=&-|C_{phn}^c|\frac{15\sqrt{\pi}}{2}\times (n_ca_s^3)^{-1/2}\nonumber\\&\times&\rho^2\text{sech}^4\bigg(\sqrt{\frac{\pi^2}{8}(\rho-1)}\bigg).
\end{eqnarray}
where $C_{phn}^c=-\frac{A\pi^2\hbar v_c}{480L^4}$ is the primary quantum Casimir force (Eqn.(7)) for the critical density.

\begin{figure}
\includegraphics{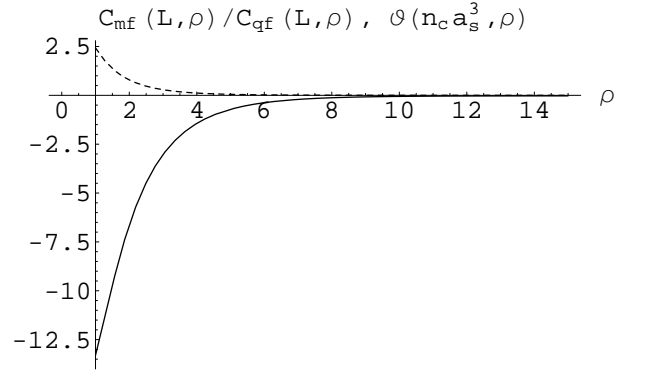}
\caption {The continuous line follows the Eqn.(32), and represents the total Casimir force in units of $|C_{phn}^c|(n_ca_s^3)^{-1/2}$. The dotted line represents the ratio of the mean field (Eqn.(28)) and fluctuations (Eqn.(31)) parts of the Casimir force in units of $10^3$. Here we set $L=10^{-7}\text{m}$ and $n_ca_s^3=4.0117\times10^{-5}$.}
\label{fig:}
\end{figure}

\subsection{Quantum fluctuations part of the Casimir force on a BEC}
Let us now go back to the Eqn.(6) whose second term ($\varepsilon_0(L,n)$) is relevant for quantum vacuum fluctuations as well as for the quantum Casimir force. Expanding $\epsilon(p_\perp,j)$ in terms of $p=\sqrt{p_\perp^2+\frac{\pi^2\hbar^2j^2}{L^2}}$, we can write the second term as
\begin{eqnarray}
\varepsilon_0(L,n)&=&\frac{1}{2}\sum_{{\bf p}{_\perp,j}}\bigg(\sqrt{\frac{gn}{m}}p\big(1-\frac{p}{\sqrt{4mgn}}+\frac{p^2}{8mgn}\nonumber\\&-&\frac{p^4}{128m^2g^2n^2}+...\big)-gn\bigg).
\end{eqnarray}
Now, replacing the sum over ${\bf p}_\perp$ by an integration from zero to an ultraviolet cutoff ($\varLambda$), we recast the Eqn.(29) as
\begin{eqnarray}
\varepsilon_0(L,n)&=&\frac{1}{2}\sum_{j=1}^\infty\frac{2\pi A}{(2\pi\hbar)^2}\bigg(\sqrt{\frac{gn}{m}}\bigg(\frac{\varLambda^3-(\pi\hbar j/L)^3}{3}\nonumber\\&-&\frac{\varLambda^4-(\pi\hbar j/L)^4}{4\sqrt{4mgn}}+\frac{\varLambda^5-(\pi\hbar j/L)^5}{40mgn}\nonumber\\&-&\frac{\varLambda^7-(\pi\hbar j/L)^7}{896(mgn)^2}+...\bigg)-gn\frac{\varLambda^2}{2}\bigg).
\end{eqnarray}
The Casimir force ($C_{qf}(L,n)=-\frac{\partial}{\partial L}\varepsilon_0(L,n)$) from the Eqn.(30) is now obtained as
$C_{qf}(L,n)=\frac{1}{2}\sqrt{\frac{gn}{m}}\frac{2\pi A}{(2\pi\hbar)^2}\big(-\frac{(\pi\hbar)^3\zeta(-3)}{L^4}+\frac{(\pi\hbar)^4\zeta(-4)}{\sqrt{4mgn}L^5}-\frac{(\pi\hbar)^5\zeta(-5)}{8mgnL^6}+\frac{(\pi\hbar)^7\zeta(-7)}{128(mgn)^2L^8}-...\big)$, which can be recast (with the regularized zeta) in terms of $\rho$ as
\begin{eqnarray}
C_{qf}(L,\rho)=-|C_{phn}^c|\rho^{1/2}\bigg(1-\frac{5}{42\rho}-\frac{1}{64\rho^2}-...\bigg).
\end{eqnarray}
The first term of the Eqn.(31) is the dominating term for the quantum fluctuations part of the Casimir force as already obtained in the Eqn.(7) and in the Refs.\cite{25,roberts}.

\subsection{Total Casimir force on a BEC}

The total Casimir force ($C(L,\rho)$) is the sum of the right hand sides of the Eqns.(28) and (31), and can be given in units of $|C_{phn}^c|(n_ca_s^3)^{-1/2}$ by
\begin{eqnarray}
\vartheta(n_ca_s^3,\rho)&=&-\bigg(\frac{15\sqrt{\pi}}{2}\rho^2\text{sech}^4\bigg(\sqrt{\frac{\pi^2}{8}(\rho-1)}\bigg)\nonumber\\&+&\sqrt{(n_ca_s^3)}\rho^{1/2}\big(1-\frac{5}{42\rho}-\frac{1}{64\rho^2}-...\big)\bigg),\nonumber\\
\end{eqnarray}
where $n_ca_s^3$ is the parameter which determines the diluteness of the condensate. Diluteness is one of the necessary conditions for achieving BEC. For the diluteness we must have $n_ca_s^3\ll 1.$ For $^{23}$Na atoms ($a_s=19.1\text{a}_0$ \cite{35}), and for $L=10^{-7}~\text{m}$, we have $n_c=3.8853\times10^{22}/\text{m}^3$ and $n_ca_s^3=4.0117\times10^{-5}$. It is to be mentioned that, $n_c=3.8853\times10^{22}/\text{m}^3$ is achievable in the 3D harmonic traps of the Bose-Einstein condensation experiments \cite{36}. For $n_ca_s^3=4.0117\times10^{-5}$, we plot the right hand side of the Eqn.(32) in the FIG. 3 with respect to the `unitless density'.

From the Eqn.(32), we get the total Casimir force as $C(L,\rho)=|C_{phn}^c|(n_ca_s^3)^{-1/2}\times\vartheta(n_ca_s^3,\rho)$. With the above parameters, and with $A=10^{-6}~\text{m}^2$, we have $|C_{phn}^c|=1.318626\times10^{-15}~\text{N}$, and $C(10^{-7}~\text{m},1)=-2.7687 \times10^{-12}~\text{N}$ which is certainly a measurable \cite{2,3} quantity.

From the Eqns.(28) and (31) we can also compare the mean field and quantum fluctuations' contributions to the Casimir force. We also plot their ratio (in units of $10^3$) in the FIG. 3. For the above parameters, the mean field part is $2425$ times stronger than quantum fluctuations part at $\rho=1$. However, at higher density ($\rho>10.96$) the fluctuation part dominates over the mean field part.

\section{Conclusions}
This paper describes the theoretical calculations on the Casimir force (as well as pressure) exerted by a self-interacting Bose-Einstein condensate on planar boundaries that confine the condensate. The calculations have been performed within the quadratic fluctuations over mean field level.

The mean field part of the Casimir force is the consequence of the inhomogeneity of the condensate. For high density $\rho\gtrsim 10$, the condensate between the two plates becomes essentially homogeneous, and the fluctuations part dominates over the mean field part.

All our calculations or results are valid only for the repulsive interaction ($a_s>0$). For $a_s<0$, the condensate becomes unstable beyond a critical number of particles \cite{biswas}. For $T\rightarrow 0$, Casimir force (in the Eqns.(7) and (27)) becomes zero \cite{28,29} only in the non-interacting limit ($a_s\rightarrow 0$), otherwise it is nonzero, attractive and experimentally measurable.

For the calculation of the Casimir force, it is necessary to change the system size at constant chemical potential by allowing particle exchange between the condensate and a particle reservoir. This is certainly convenient for the calculation, but difficult for an experiment. The Casimir force can be measured in this way only above the critical density of the atoms.

The nature of the Casimir force on the BEC (as shown in the FIG. 3), is similar to that observed \cite{15} for the $^4$He film above its lambda point.

For the consideration that the condensate is confined between two infinitely large parallel plates along the $x-y$ plane, the equation of motion of the condensate is relevant only in the $z$ direction. For the same reason, we have solved the 1D nonlinear equation (Eqn.(8)). But, practically the plates are of finite area. If $A$ be small, the equation of motion of the condensate would be relevant in the $x,y$ and $z$ directions. Calculation of the Casimir force for the plates of small area would of course be an interesting problem. In this situation, we need to solve the Eqn.(8) by replacing $\phi(z)$ by $\phi({\bf r})$. This replacement would make the Eqn.(8) a 3D nonlinear equation, which is very difficult to solve due to the fact that separation of variable technique can not be applied to the nonlinear problems. Hence, the 3D calculation for the same problem is difficult but interesting.

For a realistic case, 3D harmonically trapped BEC is to be placed between two plates. The condensate would exert a force on the plates if the plates are kept a little away from the condensate. For this realistic geometry, the magnetic field may shield the condensate from being destroyed by the (Van der Waals) interaction between the plates and the particles. If the plates are kept very far away from the condensate, then the (bulk) force acting on the plates would be zero. So, for the realistic geometry, the force acting on the plates itself is the Casimir force. The experimentalists \cite{20} already investigated the Casimir-Polder effect by putting a single plate away from a condensate. Hence, we may expect that the experimentalists might be able to measure the Casimir force for a realistic geometry of BEC.

However, we could not get the Casimir force on a 3D harmonically trapped interacting BEC. How to get this is an open question. How to calculate the Casimir force on an inhomogeneous interacting Bose gas for $T>0$, is also an open question. The same problem with interacting fermions may also be an interesting issue.

\section{Acknowledgment}
Shyamal Biswas acknowledges the hospitality and the financial support of the ``Centre for Nonlinear Studies, Hong Kong Baptist University, Kowloon Tong, HK'' for the initial work of this paper. Dwipesh Majumder thanks CSIR (of the Government of India) for the financial support. Nabajit Chakravarty is grateful to DGM of IMD for granting study leave (vide. DGM order No. A-24036/05-E(2) dt. 02.07.2007). Nabajit Chakravarty also acknowledges the hospitality of the ``Department of Theoretical Physics, Indian Association for the Cultivation of Science, Jadavpur, Kolkata 700032, India'' during the period of the study leave.


\begin{thebibliography}{99}
\bibitem{34}All the elliptic integrals $\&$ functions have been written according to the notation of Wolfram Mathematica.
\end{thebibliography}
\end{document}